# A Systematic Review of FAIR-compliant Big Data Software Reference Architectures

João Pedro de Carvalho Castro  [ Universidade de São Paulo, Universidade Federal de Minas Gerais | *jpcarvalhocastro@ufmg.br* ]
Maria Júlia Soares De Grandi [ Universidade de São Paulo | *maju.degrandi@usp.br* ]
Cristina Dutra de Aguiar  [ Universidade de São Paulo | *cdac@icmc.usp.br* ]

*Diretoria de Tecnologia da Informação, Reitoria, Universidade Federal de Minas Gerais, Avenida Presidente Antônio Carlos, 6627, Pampulha, Belo Horizonte, MG, 31270-901, Brasil.*

**Abstract.** To meet the standards of the Open Science movement, the FAIR Principles emphasize the importance of making scientific data Findable, Accessible, Interoperable, and Reusable. Yet, creating a repository that adheres to these principles presents significant challenges. Managing large volumes of diverse research data and metadata, often generated rapidly, requires a precise approach. This necessity has led to the development of Software Reference Architectures (SRAs) to guide the implementation process for FAIR-compliant repositories. This article conducts a systematic review of research efforts focused on architectural solutions for such repositories. We detail our methodology, covering all activities undertaken in the planning and execution phases of the review. We analyze 323 references from reputable sources and expert recommendations, identifying 7 studies on general-purpose big data SRAs, 13 pipelines implementing FAIR Principles in specific contexts, and 3 FAIR-compliant big data SRAs. We provide a thorough description of their key features and assess whether the research questions posed in the planning phase were adequately addressed. Additionally, we discuss the limitations of the retrieved studies and identify tendencies and opportunities for further research.

## 1 Introduction

The concept of Open Science has emerged in the scientific community to increase collaboration between researchers across the globe. It states that every digital asset originated from research objects should be made available and usable free of charge [Medeiros et al., 2020; Maedche et al., 2024]. To standardize the development of data sharing repositories capable of adhering to the Open Science concept, the FAIR Principles have been proposed [Wilkinson et al., 2016]. The objective behind these principles resides in ensuring that the aforementioned digital assets are findable, accessible, interoperable, and reusable by both humans and machines. However, their implementation might be challenging depending on the volume, variety, and velocity of the scientific data and metadata to be shared, which is a complexity inherent to big data environments [Chen et al., 2014].

Considering this complexity and the fact that the FAIR Principles are defined in proximity to the user level, a data engineer would benefit considerably from adopting a Software Reference Architecture (SRA) during the implementation process. An SRA can be defined as an architectural framework that encapsulates the expertise on creating specific system architectures (or pipelines) within a particular domain [Nakagawa et al., 2011]. Consequently, a FAIR-compliant SRA would serve as a guiding blueprint for data engineers when constructing a big data sharing repository, effectively connecting the FAIR Principles with specific implementation details.

Given the importance of big data FAIR-compliant SRAs to the context of Open Science, we conduct a systematic review of the literature encompassing this domain. A systematic review is a rigorous method that systematically searches, selects, appraises, and synthesizes existing research studies. It aims to provide an evidence-based summary of specific research questions, following predefined criteria to minimize bias while identifying research gaps [Scannavino et al., 2017]. In the literature, the work of Davoudian and Liu [2020] surveys big data SRAs available up to the year of 2020. However, the authors do not consider the FAIR principles in their comparisons. In our work, we analyze different architectural solutions for developing big data sharing repositories capable of fulfilling the FAIR Principles. To the best of our knowledge, our work is novel since no other survey covering this specific type of analysis has been retrieved during the conduction of the systematic review.

Our article presents the following contributions:

- A systematic review of the literature that employs a reproducible methodology, laying the groundwork for future research on FAIR-compliant big data SRAs.
- A detailed description of the technical background necessary to comprehend the systematic review.
- A synthesis of architectural solutions for developing repositories in line with the FAIR Principles. We also classify them as general purpose big data SRAs, context-specific pipelines to implement FAIR-compliant repositories, and FAIR-compliant big data SRAs.
- An assessment of the research questions defined in the planning phase of the systematic review, analyzing if

they have been properly answered by the studies retrieved during the conduction phase.
- A discussion encompassing the limitations and tendencies of the analyzed solutions, as well as the research opportunities that arise from these observations.

A preliminary version of our work has been published in Castro and Aguiar [2023]. Here, we further advance in the state-of-the-art by increasing the level of detail and proposing novel analyses of the obtained results. First, we provide a detailed technical background to assist in the comprehension of the systematic review. Second, we considerably increase the level of detail in the description of the employed methodology, enhancing reproducibility. Third, we also describe the retrieved studies more thoroughly, enabling better comprehension of their content. Fourth, we propose an assessment of the research questions used in the systematic review, analyzing if they have been properly answered by the studies retrieved during the conduction phase. Finally, we identify a novel tendency and novel research opportunities in the retrieved studies.

This article is structured as follows. Section 2 describes the technical background necessary to comprehend the systematic review. Section 3 presents a historical panorama regarding big data and the FAIR Principles. Section 4 details the employed methodology, including the specification of the planning phase and the conduction phase. Section 5 presents the data synthesis and classification of the studies retrieved by the systematic review. Section 6 details the assessment of the research questions. Section 7 outlines limitations, tendencies, and research opportunities. Section 8 concludes the article.

## 2 Technical Background

In this section, we describe theoretical concepts related to FAIR-compliant big data software reference architectures. Section 2.1 describes the FAIR Principles. Section 2.2 delineates the definition of software reference architectures. Section 2.3 details the concept of Big Data.

### 2.1 The FAIR Principles

The FAIR Principles encompass four fundamental categories [Wilkinson *et al.*, 2016; Jacobsen *et al.*, 2020]: Findability, Accessibility, Interoperability, and Reusability. We detail each category and the requirements that they entail as follows.

Findability pertains to locating scientific data efficiently by leveraging metadata and unique identifiers. It ensures that data can be easily discovered and accessed by both humans and machines. The following requirements need to be fulfilled to achieve Findability in a data sharing repository:

- Data and metadata are assigned a globally unique and persistent identifier;
- Data is described with rich metadata;
- Metadata clearly and explicitly include the identifier of the data it describes;
- Data and metadata are registered or indexed in a searchable resource.

Accessibility concerns the means by which data and metadata are retrieved. It is also concerned with preserving the relevance of digital resources for future utilization, facilitating understanding of data nature and provenance. To achieve Accessibility, the following requirements must be adressed:

- Data and metadata are retrievable by their identifier using a standardized communications protocol:
  - The protocol is open, free, and universally implementable;
  - The protocol allows for an authentication and authorization procedure, where necessary.
- Metadata is accessible, even when the data is no longer available.

Interoperability encompasses the representation and association of data and metadata, aiming to achieve seamless integration and communication between different systems and datasets. It establishes guidelines for using formal knowledge representation languages and aligning vocabulary with FAIR Principles, facilitating accurate interpretation and communication. A repository must address the following requirements to be Interoperable:

- Data and metadata employ a formal, accessible, shared, and broadly applicable language for knowledge representation;
- Data and metadata use vocabularies that follow FAIR Principles;
- Data and metadata include qualified references to other data and metadata.

Reusability is paramount in maximizing the value and utility of data resources, ensuring that data and metadata remain pertinent and usable over time. By providing comprehensive descriptions, clear usage licenses, and detailed provenance information, a data sharing repository can facilitate informed decision-making regarding resource reuse, fostering collaboration and knowledge exchange across various domains. To achieve proper Reusability, the following requirements should be satisfied:

- Data and metadata are richly described with a plurality of accurate and relevant attributes:
  - Data and metadata are released with a clear and accessible data usage license;
  - Data and metadata are associated with detailed provenance;
  - Data and metadata meet domain-relevant community standards.

### 2.2 Software Reference Architectures

A software reference architecture is a high-level blueprint or model that provides a structured framework for designing, building, and implementing software systems [Clements *et al.*, 2003; Nakagawa *et al.*, 2011]. It serves as a guide for software development teams and helps ensure consistency and maintainability across different projects within an organization. At its core, an SRA defines the overall structure of a software system, outlining its key components, layers (i.e. groupings of components with aligned objectives) their interactions, and the data flow between them.

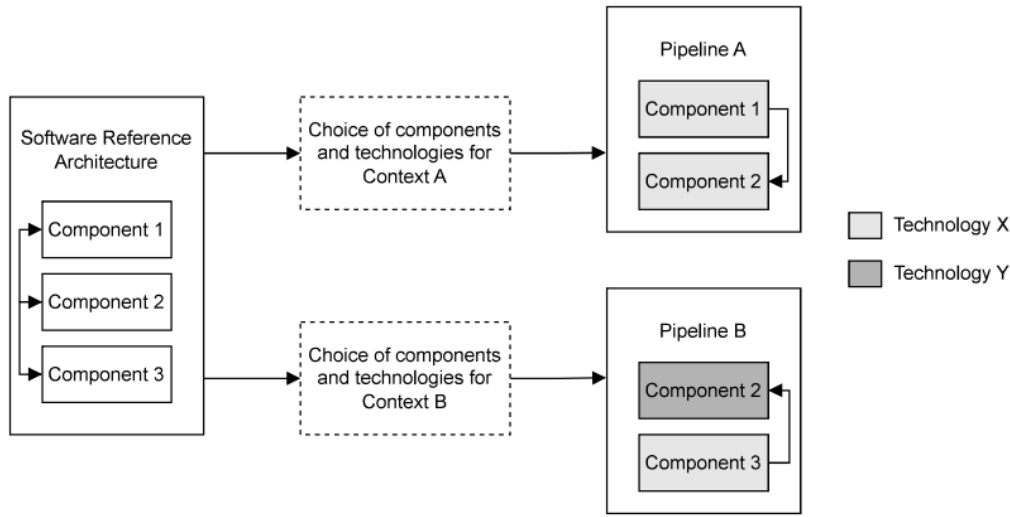

**Figure 1.** Instantiation of an SRA to two different contexts.

By being defined at a high level of abstraction, these architectures are context and technology independent. They can be instantiated as concrete workflows (or pipelines) in different contexts, each employing a different subset of technologies [Angelov *et al.*, 2012; Martinez-Fernandez *et al.*, 2015]. During instantiation, developers make concrete decisions about how the abstract components and interactions in the SRA will be realized in the specific context. This encompasses choosing programming languages, frameworks, and third-party libraries. Not all components from the SRA need to be instantiated in the pipeline, providing flexibility to developers to tailor the architecture to the unique requirements of each context.

Figure 1 depicts the process of instantiating an SRA to two different contexts. In this figure, the SRA is composed of three components that can send and receive data between themselves. This SRA is then instantiated to two different contexts, generating two different pipelines (A and B). In Pipeline A, only Components 1 and 2 are instantiated, with the data flow occurring from Component 1 to Component 2. As for Pipeline B, Components 2 and 3 are instantiated, with the data flow occurring from Component 2 to Component 3. Also, Component 2 is implemented using a different technology in each pipeline. This demonstrates that it is possible to implement a pipeline by employing different components and technologies, according to the requirements imposed by the context in which it is instantiated.

## 2.3 Big Data

Big Data refers to large and complex datasets that exceed the capabilities of traditional data processing methods. It can be characterized by three essential features: variety, volume, and velocity [Chen *et al.*, 2014; Tsai *et al.*, 2015]. Volume refers to the immense amounts of data collected daily by businesses. Thanks to the emergence of cost-effective storage technologies, companies have adopted a practice of preserving all data generated from diverse systems. Furthermore, the widespread use of the internet has significantly contributed to the expansion of data collection. For instance, metadata related to user web navigation, data from sensor networks, and Global Positioning Systems (GPS) data from smartphones serve as illustrative cases.

Variety is associated with the collection of data and metadata in structured, semi-structured, and unstructured formats [Rusu *et al.*, 2013]. Structured data can be defined as data that requires a previous schema definition, such as tabular data in relational databases. Semi-structured data is not bound by a schema definition, but nonetheless contains tags or other markers to separate semantic elements and enforce hierarchies of records and fields within the data. Unstructured data, however, does not follow a predefined data model and does not have any associated semantic elements.

Velocity refers to the speed at which data is generated, processed, and analyzed in real-time. The data flow can be continuous or rapidly changing, requiring immediate and agile processing capabilities to capitalize on its value. Technologies such as stream processing and in-memory computing enable organizations to handle the high-velocity data flow, empowering them to derive actionable insights and maintain a competitive edge.

Although it is possible to define big data based solely on the three essential features, there are works in the literature that include further characteristics on its definition. For instance, the work of Dong and Srivastava [2013] also defines big data based on its veracity, which pertains to how trustworthy is the data being analyzed. Another example is the work of Sharma and Mangat [2015], which includes both veracity and value on their definition, emphasizing that data must be important to the context of the analysis. Additionally, the work of Wrembel [2017] considers both variability and visualization when defining big data. While variability pertains to changes in value or meaning in the data, visualization refers to the manner in which this data is displayed to the data consumer.

Due to these different perspectives concerning the concept of big data, there is a lack of consensus in regards to its definition. Therefore, for simplification, we choose to define big data based on its volume, variety, and velocity [Chen *et al.*, 2014]. This approach serves as a foundational framework for the current systematic review, facilitating a comprehensive examination of pertinent literature and ensuring coherence in our analysis.

## 3 Historical Panorama of FAIR Principles and Big Data

Since their proposal in 2016, the FAIR Principles have been a subject of constant debate in the literature. As of 2019, Google Scholar registered over 2,000 accumulated citations of its original article, growing to over 14,000 accumulated citations in 2024[1], as illustrated in Figure 2. However, Wilkinson et al. [2016] described these principles at a high level of abstraction, focusing more on why the principles should be adopted rather than how they should be implemented. Due to this fact, related research faced several challenges when attempting to develop FAIR-compliant repositories, resulting in incomplete or even inaccurate implementations.

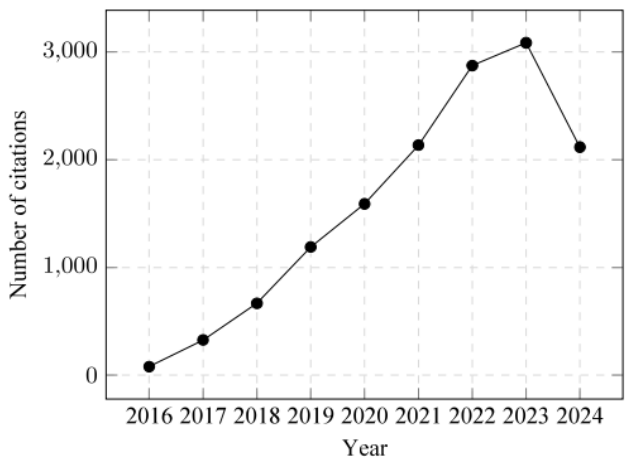

**Figure 2.** Number of citations for Wilkinson et al. [2016] over the years[1].

To address this issue, Jacobsen et al. [2020] proposed a detailed set of implementation considerations for the FAIR Principles based on extensive literature reviews. For each principle, the authors first explained its correct interpretation and then provided a comprehensive discussion regarding its implementation. The discussion encompasses the existing technologies that can be employed, the challenges that can arise, possible solutions, and examples of implementations. We synthesize this knowledge when describing the FAIR Principles in Section 2.1.

The work of Reisen et al. [2020] is also concerned with reviewing FAIR implementations. The authors reviewed studies published between 2016 and 2019, analyzing how they address the implementation of the FAIR Principles. First, the authors conducted an analysis regarding their geographic location, with the majority of implementations originating from Europe and America. The research domains of those implementations were also investigated, with over 80% of the studies pertaining to the domain of bio-science. The authors also analyzed other data pertaining to these studies, such as the distribution of research domains across geographic locations, their implementation networks, and a discussion regarding the acceptance of FAIR implementation over the years.

Since Jacobsen et al. [2020] and Reisen et al. [2020] enriched the discussion of FAIR implementations, the tendency of research after these studies is to implement FAIR-compliant repositories more accurately. Thus, our systematic review is interested in retrieving and discussing FAIR-compliant solutions that have been published after 2020.

The year 2020 is also relevant for the context of big data systems (Section 2.3) since the work of Davoudian and Liu [2020] reviewed existing SRAs (Section 2.2) available in the literature until this date. In the review, the authors describe the Lambda, Kappa, Liquid, Solid, and Bolster architectures, evaluating their big data features and comparing them with the traditional business intelligence setup. The evolution of these features is paramount for a rich analysis of big data SRAs. Therefore, we consider these architectures to be specialists' recommendations during the conduction of our systematic review. We describe and compare them in detail in Section 5.1, along with novel big data SRAs that emerged in the literature after 2020.

## 4 Methodology

A systematic review is a method for systematically summarizing research evidence. It determines that the literature review should follow a plan with phases and activities, guaranteeing reproducibility and reducing the possibility of achieving biased results. The sequence of phases and activities followed in the present systematic review is based in the work of Scannavino et al. [2017]. It is defined as follows: (i) planning, which encompasses the definition of an objective, research questions, search engines, keywords, and selection criteria; (ii) conduction, including the selection of the studies and the data synthesis; and (iii) publication of the results. This process is flexible in regards to reevaluating its phases, enabling their redefinition if necessary.

We detail the phases of our systematic review as follows. Section 4.1 describes the planning phase. Section 4.2 delineates the conduction of the systematic review.

### 4.1 Planning Phase

In this section, we describe the planning phase of the systematic review. This phase comprises several activities related to the definition of: (i) the objective of the systematic review; (ii) the research questions that will be addressed; (iii) the search engines that will be employed; (iv) the keywords and search strings derived from the research questions; and (v) the selection and quality criteria which will be employed in the assessment of the retrieved studies.

---
[1]Data extracted from Google Scholar in September 10, 2024.

### 4.1.1 Objective and Research Questions

In this activity of the planning phase, we define the objective of the systematic review as follows:

**Objective.** Identify studies that propose SRAs capable of implementing the FAIR Principles and addressing the requirements of big data environments.

From this objective, we can derive the following research questions (RQs):

- **RQ1.** Are there studies in the literature that propose implementations of the FAIR Principles for specific repository contexts?
- **RQ2.** Are there studies in the literature that propose SRAs capable of satisfying the FAIR Principles?
- **RQ3.** Are there studies in the literature that propose SRAs capable of addressing the intrinsic characteristics of big data environments?
- **RQ4.** Are there studies in the literature that propose SRAs that fulfill the FAIR Principles and address the intrinsic characteristics of big data environments?

### 4.1.2 Search Engines

The next activity is related to the definition of the search engines that we employ to conduct this systematic review. This definition is guided by their compliance with certain conditions and characteristics. The criteria observed for defining these engines are: (i) reach, indicating that the engines must return a reasonable amount of studies; (ii) novelty, specifying that the returned studies must be recent; and (iii) availability, showing that the studies must be fully accessible. We consider the following search engines:

- IEEEXplore Digital Library[2].
- ACM Digital Library[3].
- Elsevier Scopus[4].

We did not consider the DBLP search engine since most of its publications are already available in the aforementioned engines. Furthermore, its search capabilities are limited when compared with these engines [Batista *et al.*, 2018].

### 4.1.3 Keywords and Search String

To conduct the systematic review using the aforementioned search engines, we need to derive search keywords and a search string from the previously defined research questions. We define these keywords and their respective synonyms in English and Portuguese, with the exception of anglicisms. This is done so that works published in Portuguese can also be retrieved. The keywords are specified as follows:

- FAIR principles, FAIR guiding principles, *princípios FAIR*, open science, *ciência aberta*;
- Workflow, pipeline, implementation, *implementação*;
- Software reference architecture, *arquitetura de referência de software*, generic architecture, *arquitetura genérica*, SRA;
- Big data, cloud, cloud computing, *computação em nuvem*.

We do not define the synonyms for the specified keywords based on their literal definition. Instead, we consider that the same type of work can be described in various distinct ways. For instance, a pipeline that implements the FAIR Principles to a specific repository context can be described as an "open science implementation" or as a "FAIR principles workflow". Likewise, a software reference architecture for big data environments can be listed as a "generic architecture for cloud computing environments" or as a "big data SRA".

Code 1 details the search string created with the proposed keywords. We employ the logical operators *AND* and *OR* to create associations between these keywords so that the search string is able to return studies that answer the defined research questions. We employ this search string in the selected search engines, using their interface to limit the search to the article title, keywords, and abstract. We also filter articles that are not available in English or Portuguese, articles outside the field of Computer Science, and articles that have been published outside the time frame defined in the selection criteria (Section 4.1.4).

**Code 1:** Search string used in the systematic review.

```
(("FAIR principles" OR "FAIR guiding principles" OR
  "princípios FAIR" OR "open science" OR
  "ciência aberta")
  AND ("implementation" OR "implementação" OR
      "workflow" OR "pipeline"))
OR
(("software reference architecture" OR "SRA" OR
  "arquitetura de referência de software" OR
  "generic architecture" OR "arquitetura genérica")
  AND (("big data" OR "cloud" OR "cloud computing" OR
      "computação em nuvem")
      OR ("FAIR principles" OR
          "FAIR guiding principles" OR
          "princípios FAIR" OR
          "open science" OR "ciência aberta")))
```

### 4.1.4 Selection Criteria and Procedure

In this activity of the planning phase, we define the criteria for selecting the studies in our systematic review.

First, we define the time frame considered to filter the studies retrieved by the search strings. To this end, we leverage the works of Jacobsen *et al.* [2020] and Reisen *et al.* [2020]. These studies propose several recommendations for the implementation of the FAIR Principles based on a literature review conducted between the years of 2016, when the FAIR Principles were first proposed [Wilkinson *et al.*, 2016], and 2019. Since we are interested in solutions that adopt their recommendations during the definition of FAIR-compliant implementations, we disconsider works that have been published prior to the year of 2020. The same time frame can be adopted for studies that define big data SRAs, since the work of Davoudian and Liu [2020] surveys the most relevant architectures in the literature up to the year of 2020. However, we can include studies outside this time frame if they are recommended by specialists.

---

[2]IEEEXplore Digital Library: https://ieeexplore.ieee.org
[3]ACM Digital Library: https://dl.acm.org
[4]Elsevier Scopus: https://www.scopus.com

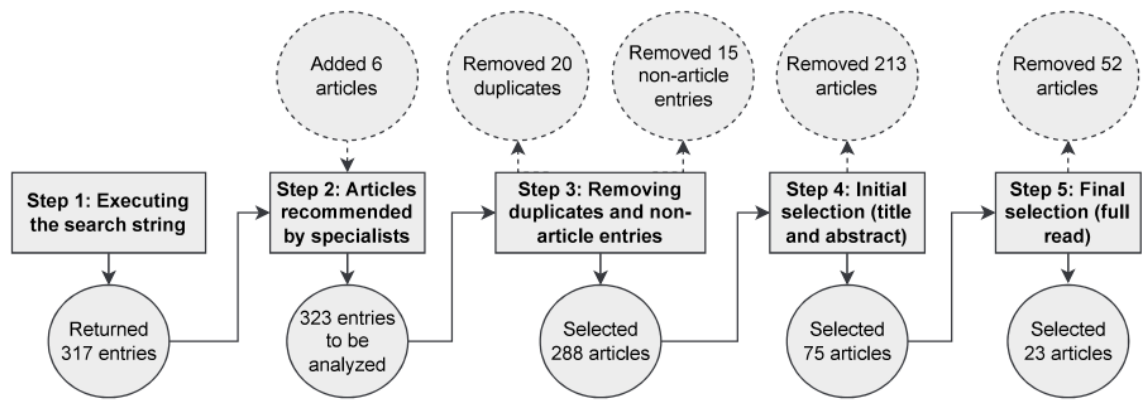

**Figure 3.** Conduction of the systematic review using the selection procedure.

With the time frame delineated, we can define the inclusion criteria. We define the following inclusion criteria for our systematic review:

- Studies that answer at least one of the defined research questions.
- Studies that have been published since the year of 2020 or recommended by specialists.
- Studies that pertain to the field of Computer Science.
- Studies that are available in English or Portuguese.
- Studies that can be accessed for free in academic environments.

Similarly, the exclusion criteria are defined as follows:

- Studies that do not answer at least one of the defined research questions.
- Studies that have been published prior to the year of 2020 and have not been recommended by specialists.
- Studies outside the field of Computer Science.
- Studies not available in English or Portuguese.
- Studies that can not be accessed for free in academic environments.
- Studies that do not meet at least one of the quality evaluation criteria.

We define the following quality evaluation criteria:

- Reputation of the journal or conference in which the study was published.
- Number of citations.
- Writing quality.

After defining the necessary criteria, we are able to delineate the selection procedure for our systematic review. This procedure is comprised of the following stages:

- **Initial Selection.** Based on the defined keywords, we identify the search string and execute it in the selected search engines. Then, we remove duplicates and non-article entries, and include studies recommended by specialists. We evaluate the studies by reading their title and abstract. If a study meets the inclusion criteria, it is chosen for the final selection. For the cases in which a study meets any exclusion criterion, it is removed from the systematic review.
- **Final Selection.** In this activity, we read the studies selected in the previous step in their entirety. The objective of this activity resides in identifying which articles still maintain the inclusion criteria. These articles are then used for the process of obtaining information.
- **Obtaining Information.** When a study complies with the inclusion criteria after the final selection, we extract its information with the objective of synthesizing its content. This process is carried out for every study that is approved in the final selection.

### 4.2 Conduction Phase

We conduct the systematic review according to the selection procedure defined in the planning phase (Section 4.1.4), as illustrated in Figure 3. In Step 1, we execute the search string (Section 4.1.3) in the chosen search engines (Section 4.1.2). The search string retrieved a total of 317 entries[5], 11 from ACM, 43 from IEEEXplore, and 263 from Scopus. Then, in Step 2, we include 6 references that have been recommended by specialists, increasing the total number of entries to 323. However, 20 of these entries are duplicates and 15 do not correspond to scientific papers. We remove these in Step 3, leaving 288 studies to be analyzed, 8 from ACM, 41 from IEEEXplore, 233 from Scopus, and 6 from specialists recommendations.

Then, we perform an initial selection (Step 4) on the obtained studies by reading their title and abstract. Of the 288 analyzed articles, 75 are approved in the initial selection since they meet the selection criteria (Section 4.1.4). Regarding the approved studies, 2 of which are from ACM, 10 from IEEEXplore, 57 from Scopus, and 6 from specialists recommendations. In the final selection (Step 5), we approved 21 studies by reading them entirely and ensuring that they still meet the selection criteria. These studies refers to 4 from IEEEXplore, 11 from Scopus, and 6 from specialists recommendations. Finally, we summarize and classify the studies into groups, as detailed in Section 5.

The amount of studies analyzed in the main steps of the systematic review is summarized in Figure 4. By analyzing the results presented in this figure, we can obtain some insights. First, with the exception of ACM, at least one study from each search engine has been approved in the final selection, which highlights the relevance of the sources. Additionally, we can observe that the amount of studies retrieved in Scopus is significantly higher than in the other search engines. This fact demonstrates the broad coverage of this search engine.

---
[5]We performed the search in April 8, 2023.

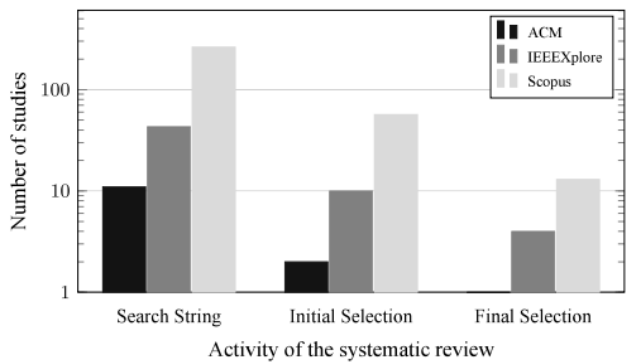

**Figure 4.** Studies selected during the systematic review, grouped by the employed search engine. A logarithmic scale is used to improve the visualization of the results.

## 5 Data Synthesis and Classification

We synthesize the content of the articles approved in the final selection and classify them into three distinct groups, based on which research questions they answer: (i) general purpose big data SRAs (Section 5.1); (ii) pipelines to implement specific FAIR repositories (Section 5.2); and (iii) FAIR-compliant big data SRAs (Section 5.3). We compare these studies based on their key characteristics. These encompass essential features for open science repositories (i.e., FAIR compliance, metadata management, source data retrieval by metadata), big data capabilities, storage of data and metadata, and being generic enough to adhere to the concept of an SRA. When relevant, we also analyze the repository context and the focus of the developed solutions.

### 5.1 Group 1: General Purpose Big Data SRAs

The research efforts allocated in this group encompass big data SRAs that have not been specifically engineered to align with the FAIR Principles. Instead, these solutions emphasize the provision of real-time analytics to support users in decision-making, without focusing on the collection and management of metadata and data provenance. We compare studies in Group 1 in Table 1 and detail them as follows.

The first work in this group describes the traditional data warehousing architecture [Chaudhuri and Dayal, 1997]. It consists of a dedicated environment for the execution of analytical queries, encompassing components such as a data warehouse, data marts, and a metadata repository. The authors describe a multidimensional and hierarchical data modeling approach in the data warehouse, along with tools for metadata management and the data warehouse itself. Also, the work covers a set of back-end and front-end tools for handling data, queries and analyses. It also includes server extensions for more efficient processing of these queries. We consider that this architecture enables big data analytics since several extensions have been proposed regarding its implementation in big data environments (e.g., the processing of star joins using parallel and distributed data processing frameworks [Brito *et al.*, 2016], a big spatial data warehouse for smart cities [Clarindo *et al.*, 2021], and an elastic data warehouse in the cloud [Ahmadi, 2023]).

The Kappa architecture [Kreps, 2014] uses a single streaming layer for big data computation. It supports batch and real-time processing through the buffering of historical data in a logging system for an extended duration. Also, this architecture provides a straightforward approach for its purpose in terms of coding, testing, and debugging.

The Lambda architecture [Kiran *et al.*, 2015] employs three streaming layers, each of which responsible for: (i) creating batch views; (ii) processing recent data into real-time views; and (iii) storing and merging the views for later consumption. This architecture represents a cost-effective approach through the use of cloud computing. It also has the capacity to handle a large volume of data.

Liquid [Fernandez *et al.*, 2015] is an SRA that aims to overcome the limitations of Kappa and Lambda by employing incremental processing instead of fully recomputing views. This architecture also considers aspects such as high availability, cost efficiency, low latency, and resource isolation. Liquid encompasses two layers. The first one, named messaging, stores data and metadata as messages. It also manages checkpoints from which this data can be partially recomputed. The second layer, called processing, performs jobs and transformations on the messages. Although the messaging layer stores metadata, Liquid does not include any implementation regarding its management.

The Solid architecture [Martínez-Prieto *et al.*, 2015] unifies heterogeneous data into a single model, defining layers for big data storage, querying, streaming, and merging of historical and runtime data. It distinguishes itself from the Lambda architecture by streamlining data management through a single-layer storage approach, abolishing the need for duplicating data into separate batch and real-time views. In addition to increasing data integration capabilities, Solid optimizes storage efficiency, especially for extensive datasets.

As for the Bolster architecture [Nadal *et al.*, 2017], the authors use a semantic layer with metadata management for data governance. This includes an ontology-based repository for input data characteristics, accompanied by a dispatcher component that determines whether the streaming data should be directed to the batch or speed layer. Also, Bolster uses metadata to simplify tasks related to defining and exploring data, achieving partial automation in data exploration and providing support in decision-making processes.

Furthermore, NeoMycelia [Ataei and Litchfield, 2021] is an SRA based on microservices and events. Each microservice has a local database with a caching component. This architecture contains several components, such as: (i) a gateway for user connection; (ii) controllers and service meshes for stream and batch data processing; (iii) an event backbone and an event archive to support the communication between microservices; (iv) a data lake that stores structured, unstructured, and semi-structured data; (v) a query controller and query engine to support query execution; and (vi) a seman-

Table 1. Comparison of general purpose big data SRAs.

| Work* | Fits the concept of an SRA | FAIR-compliant | Metadata management | Source data retrieval by metadata | Enables big data analytics | Storage of data and metadata |
|---|---|---|---|---|---|---|
| Chaudhuri and Dayal [1997] | ✓ | ✗ | ✓ | ✗ | ✓ | Same infrastructure |
| Kreps [2014] | ✓ | ✗ | ✗ | ✗ | ✓ | Does not apply |
| Kiran et al. [2015] | ✓ | ✗ | ✗ | ✗ | ✓ | Does not apply |
| Fernandez et al. [2015] | ✓ | ✗ | ✗ | ✗ | ✓ | Does not apply |
| Martínez-Prieto et al. [2015] | ✓ | ✗ | ✗ | ✗ | ✓ | Does not apply |
| Nadal et al. [2017] | ✓ | ✗ | ✓ | ✗ | ✓ | Same infrastructure |
| Ataei and Litchfield [2021] | ✓ | ✗ | ✓ | ✗ | ✓ | Same infrastructure |

* Architectures proposed prior to the year of 2020 are classified in this group due to their inclusion in the systematic review as recommendations from specialists.

tic layer, which contains a metadata management system responsible for storing metadata, preparation rules, and data evolution.

## 5.2 Group 2: Pipelines to Implement Specific FAIR Repositories

Studies classified in this group represent pipelines that implement the FAIR Principles to the context of specific data sharing repositories. Solutions that do not implement data sharing repositories are disconsidered, such as FAIRness assessment tools or workflows to conduct specific scientific experiments. We compare studies in Group 2 in Table 2 and detail them as follows.

In Assante et al. [2021], the authors propose the AGINFRA PLUS platform, which enables researchers to store, analyze, visualize, and publish agriculture and food data in accordance with open science. The platform employs an open and distributed approach, integrating technologies such as big data, web, and cloud in agriculture and food science research. The platform encourages collaboration and facilitates access to significant resources for these communities.

Pana et al. [2021] developed a pipeline that extracts data and metadata from several seismic databases available online, integrating them into a single repository stored in a PostgreSQL local database to enable analytics. The initiative aims to aid the scientific community in sharing and reproducing data by employing the FAIR principles, emphasizing the potential for its automated retrieval and utilization.

Furthermore, the work of Pestryakova et al. [2022] describes a pipeline that extracts data and metadata from COVID-19 related publications, transforming them into CovidPubGraph, a knowledge graph which, in itself, can be considered as a data sharing repository. The work also describes the complexity entailed in the process of finding organized and easily manipulable data about diseases. For that reason, the knowledge graph created follows the FAIR principles, ensuring these properties.

The Body in Numbers system [Bruha et al., 2022] encompasses the collection of health-related data and metadata taking into consideration the FAIR Principles. The authors propose a pipeline that consists of five modules to collect, annotate, analyze, interpret, and publish brain and physical data and its associated metadata. They combine physical and brain data to investigate how they integrate, aiming to enhance the quality and duration of human life. Furthermore, the authors demonstrate that the adoption of best practices, coupled with the FAIR principles, can be crucial to improve the tasks performed by the pipeline.

In Jha et al. [2022], the authors propose a pipeline to extract data and metadata from several healthcare systems. A series of Python scripts is employed to extract image features and perform data cleaning and integration, storing the result as data triples. The authors highlight that they faced several challenges related to standardization and scalability. Therefore, they emphasize how adhering to FAIR principles is essential for their proposed infrastructure.

Additionally, the work of Felikson et al. [2022] describes a cloud infrastructure that provides a common environment for numerical model computation, NASA satellite observations, and analysis workflows behind the repository of NASA's Earth Information System. The goal of the architecture resides in enabling researchers and end users to conduct their own analyses close to big data stored in the cloud. The architecture is also aimed to make it easier to access data products and information, to reproduce analyses, and to build on existing work, following the concept of Open Science.

The VODAN BR project [Borges et al., 2022] aims to collect and implement a data management infrastructure for COVID-19 hospitalized patients' cases in Brazil, according to the FAIR principles. The authors describe the proposed architecture, which covers the processes between the collection of clinical data to the publication of its metadata in the network as triplestores.

NEANIAS [Sciacca et al., 2022] is a service-oriented architecture to provide analytics for underwater, atmospheric and space related data, aiming to create a collaborative research environment. It is comprised of four high level core services to support open science lifecycles, integration with the open science cloud, artificial intelligence, and visualization. The architecture also seeks to provide infrastructure for storing, managing, transporting, and processing large volumes of data.

The work of Toulet et al. [2022] describes a pipeline to extract metadata from scientific papers and to generate knowledge from their full text, storing both as triplestores in a Virtuoso server. To this end, this project employs natural language processing, semantic web, and linked data techniques. Additionally, it develops search and visualization services for

Table 2. Comparison of pipelines to implement specific FAIR repositories.

| Work | Fits the concept of an SRA | FAIR-compliant | Metadata management | Source data retrieval by metadata | Enables big data analytics | Storage of data and metadata | Context |
|---|---|---|---|---|---|---|---|
| Assante et al. [2021] | ✗ | ✓* | ✓ | ✓ | ✗ | Same infrastructure | Agriculture and food data |
| Pana et al. [2021] | ✗ | Partially | ✓ | ✓ | ✗ | Same infrastructure | Earthquake data |
| Pestryakova et al. [2022] | ✗ | ✓ | ✓ | ✓ | ✗ | Same infrastructure | COVID-19 papers data |
| Bruha et al. [2022] | ✗ | ✓* | ✓ | ✓ | ✗ | Same infrastructure | Health and brain data |
| Jha et al. [2022] | ✗ | Partially | ✓ | ✓ | ✗ | Same infrastructure | Oncology data |
| Felikson et al. [2022] | ✗ | ✓* | ✓ | ✓ | ✓ | Same infrastructure | Earth data from NASA |
| Borges et al. [2022] | ✗ | ✓* | ✓ | ✓ | ✗ | Same infrastructure | COVID-19 patients data |
| Sciacca et al. [2022] | ✗ | ✓* | ✓ | ✓ | ✓ | Same infrastructure | Underwater, atmospheric, and space data |
| Toulet et al. [2022] | ✗ | ✓* | ✓ | ✓ | ✗ | Same infrastructure | Textual data from papers |
| Schwagereit et al. [2022] | ✗ | ✓ | ✓ | ✓ | ✗ | Same infrastructure | In vivo data |
| Deng et al. [2022] | ✗ | ✓* | ✓ | ✓ | ✗ | Separate infrastructures | Immunology data |
| Rueda-Ruiz et al. [2022] | ✗ | ✓* | ✓ | ✓ | ✓ | Same infrastructure | LiDAR data |
| Lehmann et al. [2023] | ✗ | ✓* | ✓ | ✓ | ✓ | Same infrastructure | Sensor data |

*The authors state that the pipeline is FAIR-compliant, however no details are given on how it fulfills each FAIR principle.

managing the generated results, aiming to assist researchers in overcoming challenges when searching for relevant articles in their research. This is especially relevant as searching based only on keywords often fails to discover interesting associations.

In Schwagereit et al. [2022], the authors describe FISH, a platform to share in vivo data according to the FAIR Principles. It consists of multiple management and storage components available through microservices. The platform was conceived based on the challenges related to the exposure and integration of data and metadata in industrial production environments, which requires the management of a significant amount of data and systems. With FISH, the authors aim to achieve machine-actionable interoperability and semantic awareness.

Deng et al. [2022] propose ImmuneData, a platform that extracts and integrates the metadata of different immunology databases into a single repository, which stores this metadata in a unified metadata model proper for biomedical data. Therefore, users can use a single engine to query the metadata to retrieve the source data objects through a web interface and a search engine that converts user queries into ontology terms. The engine is also expanded by natural language processing technologies, ensuring results that encompass all synonyms of a given concept.

In Rueda-Ruiz et al. [2022], the authors propose a general specification for cloud repositories to store large scale LiDAR data. It consists of a conceptual data model implemented on MongoDB and an API to handle requests. The authors also highlight the impact of using the project in public organizations, collaborative repositories, and companies, requiring a centralized repository to handle their data.

Finally, the work of Lehmann et al. [2023] proposes an architecture that merges the concepts of Research Data Management (RDM) based on the FAIR Principles with the concept of a digital twin, which is the virtual counterpart of a physical sensor. The architecture collects sensor data and metadata and sends them to a layer called RDM Core Space, where the data is stored in its raw format and the metadata is stored as a knowledge graph. The sensor data and metadata can then be used by smart applications through a messaging broker. Furthermore, the authors emphasize the importance of integrating digital twins and the FAIR Principles to overcome challenges in data management within research institutes that extensively utilize this type of equipment.

## 5.3 Group 3: FAIR-compliant Big Data SRAs

This group of studies encompasses architectural frameworks that are context and technology independent, thus being able to adhere to the concept of an SRA (Section 2.2). Additionally, studies in this group are concerned with the requirements imposed by the FAIR Principles and employ solutions to handle the three essential features of big data environments (i.e., volume, variety, and velocity), such as parallel and distributed data processing and cloud computing technologies. We compare studies in Group 3 in Table 3 and detail them as follows.

Table 3. Comparison of FAIR-compliant big data SRAs.

| Work | Fits the concept of an SRA | FAIR-compliant | Metadata management | Source data retrieval by metadata | Enables big data analytics | Storage of data and metadata | Focus |
|---|---|---|---|---|---|---|---|
| Castro et al. [2022a] | ✓ | ✓ | ✓ | ✓ | ✓ | Separate infrastructures | Flexibility, data ownership |
| Castro et al. [2022b] | ✓ | ✓ | ✓ | ✓ | ✓ | Same infrastructure | Performance, simplification for providers |
| Vazquez et al. [2022] | ✓ | Partially | ✓ | ✓ | ✓ | Same infrastructure | Metadata quality assurance |

The work of Castro et al. [2022a] proposes BigFAIR, a FAIR-compliant SRA to store, process, and query scientific data and metadata. This architecture is comprised of several layers organized in two separate infrastructures: (i) local, encompassing the local environments of the data providers, where the source data objects remain stored; and (ii) repository, encompassing big data technologies for centralized metadata storage, data and metadata processing, and ad-hoc data anonymization. Metadata is stored either in the Metadata Lake in its raw format, or in the Metadata Warehouse after undergoing transformations, which ensures metadata persistence even when the associated source data objects no longer exist. The authors detail the compliance of each layer with each FAIR Principle. By taking advantage of the existing local infrastructures of data providers, this architecture is able to support data ownership and increase flexibility. The authors validate the proposed architecture by conducting a case study with a real-world dataset of Brazilian COVID-19 patients, demonstrating its adherence to the Open Science movement and its support for big data analytics.

CloudFAIR [Castro et al., 2022b] is a FAIR-compliant SRA that handles scientific data and its associated metadata in a single cloud infrastructure. The authors claim that this unification unburdens data providers in regards to the management of a local infrastructure and also improves performance. By being an extension of BigFAIR, CloudFAIR inherits its adherence to the Open Science movement and its full compliance with the FAIR Principles, as well as the storage of transformed metadata in a Metadata Warehouse to guarantee its persistence. However, instead of using a Metadata Lake to store only metadata in its raw format, CloudFAIR uses a multi-tiered Data Lake that stores two copies of the source data objects: a fully anonymized copy and an encrypted copy without anonymization. The authors conduct a performance evaluation that proves that this storage strategy, along with the other features of CloudFAIR, improve performance in up to 75.95% when compared to BigFAIR.

Finally, the work of Vazquez et al. [2022] proposes GADDS, a generic platform that stores research data and metadata in the cloud, achieving partial compliance with the FAIR Principles. The designed architecture is comprised of a user interface, a cloud storage infrastructure, and a blockchain environment. In this blockchain setting, metadata is cryptographically stored, ensuring quality control since every entry is validated by every node in the network in a decentralized manner. A version control software is also employed to track changes in the metadata and guarantee its persistence even when the associated source data object is excluded. However, GADDS is unable to track changes in the data objects. Thus, if a data object changes, the older versions of its associated metadata will point to the novel version of the data object. These data objects are stored by GADDS in an object storage in the cloud that enables data replication into multiple nodes, allowing parallel and distributed data processing. The authors validate GADDS with a case study related to tissue engineering and discuss its FAIR compliance at a high level.

# 6 Research Questions Assessment

After synthesizing the content of the articles approved in the final selection (Section 5), we can assess if the research questions proposed in the planning phase of the systematic review (Section 4.1.1) have been properly answered. We analyze each proposed research question as follows.

**RQ1. Are there studies in the literature that propose implementations of the FAIR Principles for specific repository contexts?** The first research question aims to identify FAIR implementations in the literature that are not generic enough to adhere to the concept of an SRA. Rather, these studies aim to implement pipelines that build specific FAIR-compliant repositories. During the conduction of our systematic review, we were able to identify 13 articles that answer this research question. These solutions implement FAIR-compliant repositories to the most diverse contexts, ranging from health-related implementations to earth sciences.

**RQ2. Are there studies in the literature that propose SRAs capable of satisfying the FAIR Principles?** The objective of this research question resides in retrieving studies that propose FAIR-compliant SRAs. In the scope of RQ2, there is no restriction in regards to big data analytics: we aim to retrieve even those SRAs that do not consider big data environments in their conception. We have identified 3 studies that are able to answer RQ2. However, all of these solutions are also aware of the intrinsic characteristics of big data environments; thus, they are also able to answer RQ4. Although this intersection might render the answer of RQ2 redundant, this might not prove to be true in future updates of the systematic review. Therefore, it is of paramount importance for RQ2 to remain in its planning phase. By doing so, if a FAIR-compliant SRA that does not employ big data solutions emerges in the future, it can be considered among the analyzed solutions.

**RQ3. Are there studies in the literature that propose SRAs capable of addressing the intrinsic characteristics**

Table 4. Studies that answer each research question.

| Research question | Studies that answer the research question | Total |
|---|---|---|
| RQ1 | Assante *et al.* [2021], Pana *et al.* [2021], Pestryakova *et al.* [2022], Bruha *et al.* [2022], Jha *et al.* [2022], Felikson *et al.* [2022], Borges *et al.* [2022], Sciacca *et al.* [2022], Toulet *et al.* [2022], Schwagereit *et al.* [2022], Deng *et al.* [2022], Rueda-Ruiz *et al.* [2022], Lehmann *et al.* [2023] | 13 |
| RQ2 | Castro *et al.* [2022a], Castro *et al.* [2022b], Vazquez *et al.* [2022] | 3 |
| RQ3 | Chaudhuri and Dayal [1997], Kreps [2014], Kiran *et al.* [2015], Fernandez *et al.* [2015], Martínez-Prieto *et al.* [2015], Nadal *et al.* [2017], Ataei and Litchfield [2021] | 7 |
| RQ4 | Castro *et al.* [2022a], Castro *et al.* [2022b], Vazquez *et al.* [2022] | 3 |

of big data environments? This research question intends to analyze the panorama of general purpose big data SRAs in the literature. Therefore, we are looking for architectures that were not concerned with satisfying the FAIR principles during their conception. Instead, their proposal focused on providing analytics for a considerable volume and variety of data produced at a high velocity. A total of 7 studies that answer RQ3 have been identified, all of which are generic enough to be considered SRAs while also being concerned with big data constraints.

**RQ4. Are there studies in the literature that propose SRAs that fulfill the FAIR Principles and address the intrinsic characteristics of big data environments?** The last research question is interested in retrieving solutions that combine all of the previously stated requirements: (i) adhering to the concept of an SRA; (ii) fulfilling the FAIR principles; and (iii) being concerned with the constraints inherent to big data environments. Solutions that answer this research question are the primordial object of interest of this systematic review. They enable data engineers to build modern Open Science repositories, assisting these professionals in complying with the FAIR Principles while also being prepared to handle big data and metadata. Our systematic review was able to identify 3 studies that answer RQ4. All of the identified studies fit the concept of an SRA and are aware of the intrinsic characteristics of big data environments. However, only two of the analyzed solutions are able to achieve full FAIR compliance. Also, each solution is focused on a different user profile, such as those that prefer flexibility, performance, or metadata quality control. Thus, a data engineer has to select which solution better aligns with the context of the repository being developed.

Table 4 summarizes the studies that answer each research question. In this table, we can analyze that the majority of studies retrieved in the systematic review answer RQ1, while RQ2 and RQ4 are answered by the minority. This can indicate that the scientific community in general is more interested in developing implementations of the FAIR Principles to specific contexts, rather than developing architectures that are able to assist data engineers to implement the FAIR Principles in any given context. It also highlights that there should be more research opportunities in the latter, since having a smaller amount of published studies might indicate that it remains a relatively underexplored research field.

## 7 Discussion

In this section, we introduce discussions regarding the studies approved in the final selection of the systematic review. We highlight limitations (Section 7.1), tendencies (Section 7.2), and research opportunities (Section 7.3).

### 7.1 Limitations

The objective of this systematic review resides in retrieving studies that propose SRAs capable of implementing the FAIR Principles while also addressing the intrinsic characteristics of big data environments. However, the studies synthesized in Section 5 face some limitations in this regard, described as follows and depicted in Figure 5.

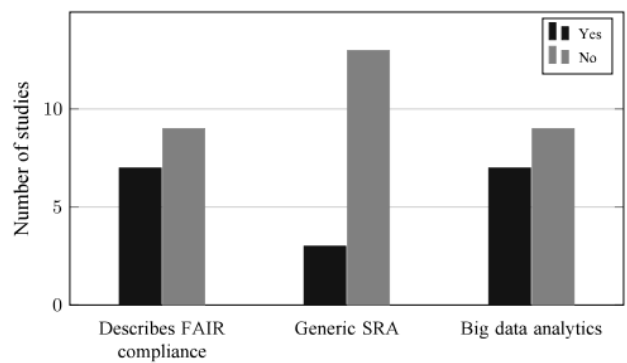

**Figure 5.** Main limitations of studies in Groups 2 and 3.

Studies in Group 1 (Section 5.1) inherently diverge from the concept of a FAIR-compliant SRA in regards to their purpose. First, these SRAs fall short in meeting the requirements set by the FAIR Principles due to their limited capabilities for retrieving source data objects based on metadata and for keeping metadata alive even when the associated data objects are no longer available. Additionally, these architectures do not employ a specific component for storing metadata about the metadata. They either do not store this content or store all the metadata in the same component, compromising the richness and performance of metadata analyses.

Regarding Group 2 (Section 5.2), none of the reviewed studies propose architectures generic enough to fit the concept of an SRA. Rather, they propose pipelines that are specific to implement a FAIR-compliant repository in a given context. Furthermore, the majority of these studies fail to clarify which of the FAIR Principles are satisfied by their solutions. Only the studies of Pestryakova *et al.* [2022] and Schwagereit *et al.* [2022] detail how their full FAIR compli-ance is achieved, while the studies of Pana *et al.* [2021] and Jha *et al.* [2022] clarify their partial compliance. This lack of information gives rise to substantial concerns, such as making sure that all principles are fulfilled and identifying which part of the solution is responsible for implementing each principle. Studies in this group are also limited in regards to their big data capabilities. Only the studies of Felikson *et al.* [2022], Sciacca *et al.* [2022], Rueda-Ruiz *et al.* [2022], and Lehmann *et al.* [2023] employ big data technologies in the construction of their solutions. Although enabling big data analytics is not a requirement imposed by the FAIR Principles, it is of significant importance to support the decision-making process. It not only allows data consumers to perform different types of analyses on the stored data and metadata, but also contributes to an increase in the overall performance of the repository.

The studies classified in Group 3 (Section 5.3) also present some limitations. For instance, GADDS [Vazquez *et al.*, 2022] is unable to achieve full compliance with the FAIR Principles. By storing metadata in a blockchain environment, it can only be exposed to members inside the network. This hinders the general public unable to access the content of a repository implemented by GADDS, compromising findability and accessibility. Also, this SRA does not implement global unique identifiers, further impacting on its FAIR compliance. Another limitation of the studies in this group is related to their focus. For instance, CloudFAIR [Castro *et al.*, 2022b] uses a single cloud infrastructure to store data and metadata to improve performance and to unburden data providers. However, by doing so it relinquishes support to data ownership and flexibility, key features of BigFAIR [Castro *et al.*, 2022a]. This can be a problem in situations in which the repository is required to comply with different data protection regulations, or to implement specific security policies. The reverse situation is also true: BigFAIR forsakes performance and simplification for data providers in order to support data ownership and obtain flexibility. Also, by not using a blockchain environment like GADDS, both BigFAIR and CloudFAIR relinquish decentralized metadata control, negatively impacting on the quality assurance of stored metadata.

## 7.2 Tendencies

After analyzing the data synthesis obtained in this systematic review (Section 5), we can identify some tendencies in the described studies. For instance, applying the FAIR Principles to health-related data is a common occurrence. Between the studies of Group 2 (Section 5.2), the following adhere to this context: (i) Pestryakova *et al.* [2022], with data from COVID-19 research publications; (ii) Bruha *et al.* [2022], using health and brain data; (iii) Jha *et al.* [2022], employing oncology related data; (iv) Borges *et al.* [2022], using data from COVID-19 patients; and (v) Deng *et al.* [2022], leveraging immunology data. Additionally, both BigFAIR [Castro *et al.*, 2022a] and CloudFAIR [Castro *et al.*, 2022b] use COVID-19 patients data in their experiments, whereas GADDS [Vazquez *et al.*, 2022] employs fiber cell tissue research data during its instantiation. Another commonly addressed context is that of earth-related data, being covered by the repositories of Pana *et al.* [2021], Felikson *et al.* [2022], Sciacca *et al.* [2022], and Rueda-Ruiz *et al.* [2022]. The proportion of studies per data context is illustrated in Figure 6.

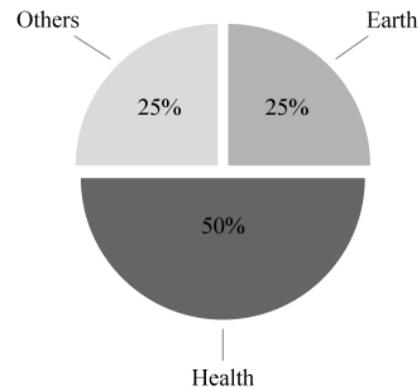

**Figure 6.** Main data contexts of studies in Groups 2 and 3.

Another observed tendency is the adoption of microservices for the construction of architectures and pipelines, which is a paradigm to deploy applications as a collection of events that are inherently independent. This strategy is observed in NeoMycelia [Ataei and Litchfield, 2021], the latest general purpose big data SRA available in the literature, and in the FISH platform [Schwagereit *et al.*, 2022], a pipeline to implement an in vivo data repository in accordance with the FAIR Principles.

In a chronological perspective, we can observe a tendency regarding the adoption of big data technologies in the proposed solutions. If we consider the studies in Groups 2 and 3, none of those proposed in 2021 (i.e. 0%) enables big data analytics in their conception. This panorama changes considerably in 2022, with about 46% of these studies being concerned with the intrinsic characteristics of big data. The work published in 2023 and retrieved by the systematic review enables big data analytics, which entails in a 100% statistic for this year. Figure 7 summarizes this chronological tendency.

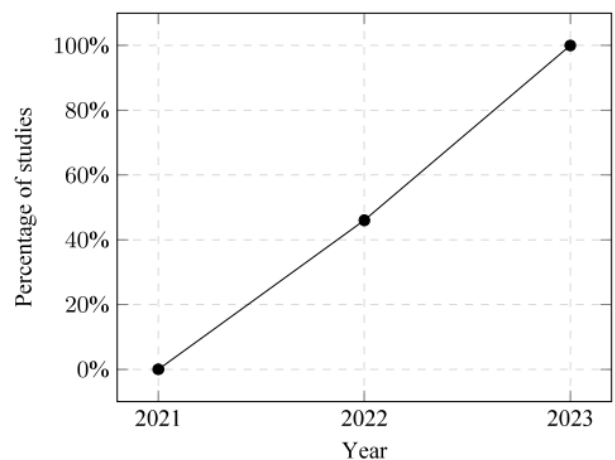

**Figure 7.** Percentage of studies from Groups 2 and 3 that enable big data analytics over the years.

Finally, the employment of the same infrastructure for the management of data and metadata is another detected tendency. The majority of the surveyed solutions use this strategy, with only a few exeptions [Deng *et al.*, 2022; Castro *et al.*, 2022a]. According to the experiments conducted in Castro *et al.* [2022b], a possible reason for the occurrence of this tendency is the improvement of query performance when storing data and metadata in the same infrastructure. This strategy also unburdens data providers in regards to maintaining a local repository to store scientific data, while overlooking data ownership and flexibility.

### 7.3 Research Opportunities

The aforementioned limitations and tendencies give rise to opportunities to conduct innovative research. For instance, the development of a novel FAIR-compliant SRA capable of unifying the advantages of BigFAIR [Castro *et al.*, 2022a], CloudFAIR [Castro *et al.*, 2022b], and GADDS [Vazquez *et al.*, 2022] would considerably benefit the scientific community. However, leveraging flexibility, data ownership, performance, simplification for data providers, and metadata quality assurance in a single architecture is challenging. A possible solution is developing this SRA in multiple modules, each prioritizing one of the aforementioned characteristics. These modules can then be instantiated depending on the requirements imposed by the data providers and consumers.

Another opportunity that arises from the previously identified tendencies is the development of a FAIR-compliant SRA using the strategy of microservices. This strategy has successfully been employed by Ataei and Litchfield [2021] for developing a general purpose big data SRA and by the work of Schwagereit *et al.* [2022] for implementing a FAIR-compliant repository. However, we have not identified a solution that merges these research fields during the conduction of the systematic review, representing a gap that can be further explored by the scientific community.

We can also observe a research opportunity in regards to the evolution of existing FAIR-compliant solutions to enable big data analytics. While most of the studies from Group 2 do not support big data analytics, we have observed a tendency to provide this type of support in more recent solutions. Thus, selecting relevant implementations from Group 2 and extending them with big data technologies could significantly benefit the scientific community. This would enable not only performance gains, but also the ability to generate different analytics and knowledge from existing data and metadata.

Furthermore, the creation of a benchmark to compare FAIR-compliant SRAs can also be perceived as a research opportunity. A benchmark can be defined as a standardized method for comparison that adheres to four criteria [Bog, 2014]. First, it should offer relevant metrics facilitating comparisons among systems within a particular domain. Second, it must demonstrate portability, allowing easy adaptation to diverse systems and architectures. Third, scalability is crucial, enabling the assessment of system performance across varying data volumes. Fourth, results should be presented in a clear and easily interpretable manner. Additionally, factors such as repeatability, fairness, verifiability, and cost-effectiveness are equally important considerations for a benchmark. Thus, creating a benchmark in the context of FAIR-compliant SRAs would enable a comparison of these architectures in a standardized manner, considerably benefiting the scientific community.

## 8 Conclusions and Future Work

In this article, we presented a systematic review of the literature that identified architectural solutions capable of implementing the FAIR Principles and addressing the intrinsic characteristics of big data environments. We described the necessary technical background and a historical panorama of the area to assist in the comprehension of the systematic review. We also detailed its methodology and conduction, enabling reproducibility. Moreover, we introduced a data synthesis of the selected studies, as well as their classification in three distinct groups. We then assessed the research questions based on the studies retrieved during the conduction of the systematic review. We also identified the limitations of these solutions, deriving tendencies and research opportunities.

Future work consists on analyzing a broader scope of studies by considering the snowball technique. With more studies being analyzed, we intend to identify new limitations, tendencies, and research opportunities. We also intend to provide a detailed evolution of the solutions across the years. Additionally, since our systematic review includes only works published after 2020, we also intend to thoroughly examine earlier reviews in this field. By comparing these prior reviews with our findings, we aim to identify any persistent issues and limitations, and assess whether current solutions have effectively addressed these problems. Furthermore, we plan on providing future updates for this systematic review and to explore the identified research opportunities. For instance, we aim to propose novel architectural solutions for the implementation of the FAIR Principles in big data environments. Another opportunity that we intend to explore is the evolution of existing FAIR-compliant implementations to enable big data analytics.

## Acknowledgements

This work was supported by the São Paulo Research Foundation (FAPESP), the Brazilian Federal Research Agency (CNPq), and the Coordenação de Aperfeiçoamento de Pessoal de Nível Superior, Brasil (CAPES).

## Funding

This research was funded by CAPES (Finance Code 001) and FAPESP (Grant #2018/22277-8).

## Authors' Contributions

João P. C. Castro is responsible for the conception and conduction of this systematic review, as well as writing the majority of the manuscript. Maria J. S. De Grandi and Cristina D. Aguiar have contributed in writing and reviewing the manuscript.

## Competing interests

The authors declare that they have no known competing financial interests or personal relationships that could have appeared to influence the work reported in this article.

## Availability of data and materials

Data will be made available on request.

# References

Ahmadi, S. (2023). Elastic data warehousing: Adapting to fluctuating workloads with cloud-native technologies. *Journal of Knowledge Learning and Science Technology ISSN: 2959-6386 (Online)*, 2(3):282–301. DOI: 10.60087/jklst.vol2.n3.p301.

Angelov, S., Grefen, P., and Greefhorst, D. (2012). A framework for analysis and design of software reference architectures. *Information and Software Technology*, 54(4):417–431. DOI: 10.1016/j.infsof.2011.11.009.

Assante, M., Boizet, A., Candela, L., Castelli, D., Cirillo, R., Coro, G., Fernandez, E., Filter, M., Frosini, L., Kakaletris, G., et al. (2021). Realising a science gateway for the agrifood: the AGINFRA PLUS experience. In *CEUR Workshop Proc.*

Ataei, P. and Litchfield, A. (2021). NeoMycelia: A software reference architecture for big data systems. In *Proc. APSEC*, pages 452–462. DOI: 10.1109/APSEC53868.2021.00052.

Batista, N. A., Sousa, G. A., Brandão, M. A., da Silva, A. P. C., and Moro, M. M. (2018). Tie strength metrics to rank pairs of developers from github. *Journal of Information and Data Management*, 9(1):69–69. DOI: 10.5753/jidm.2018.1637.

Bog, A. (2014). *Benchmarking Transaction and Analytical Processing Systems: The Creation of a Mixed Workload Benchmark and its Application*. In-Memory Data Management Research. Springer Berlin Heidelberg, Berlin, Heidelberg.

Borges, V., de Oliveira, N. Q., Rodrigues, H., Campos, M., and Lopes, G. (2022). A platform to generate FAIR data for COVID-19 clinical research in Brazil. In *Proc. ICEIS*, pages 218–225. DOI: 10.5220/0011066800003179.

Brito, J. J., Mosqueiro, T., Ciferri, R. R., and Ciferri, C. D. A. (2016). Faster cloud star joins with reduced disk spill and network communication. *Procedia Computer Science*, 80:74–85. DOI: 10.1016/j.procs.2016.05.299.

Bruha, P., Mouček, R., Salamon, J., and Vacek, V. (2022). Workflow for health-related and brain data lifecycle. *Frontiers in Digital Health*, 4. DOI: 10.3389/fdgth.2022.1025086.

Castro, J. P. C. and Aguiar, C. D. (2023). Big data architectures for FAIR-compliant repositories: A systematic review. In *Anais do XXXVIII Simpósio Brasileiro de Bancos de Dados*, pages 76–88. DOI: 10.5753/sbbd.2023.232494.

Castro, J. P. C., Romero, L. M., Carniel, A. C., and Aguiar, C. D. (2022a). FAIR Principles and Big Data: A software reference architecture for Open Science. In *Proc. ICEIS*, pages 27–38. DOI: 10.5220/0011045500003179.

Castro, J. P. C., Romero, L. M., Carniel, A. C., and Aguiar, C. D. (2022b). Open Science in the cloud: The CloudFAIR architecture for FAIR-compliant repositories. In *Proc. ADBIS*, pages 56–66. DOI: 10.1007/978-3-031-15743-1_6.

Chaudhuri, S. and Dayal, U. (1997). An overview of data warehousing and OLAP technology. *SIGMOD Record*, 26(1):65–74. DOI: 10.1145/248603.248616.

Chen, M., Mao, S., and Liu, Y. (2014). Big data: A survey. *Mobile Networks and Applications*, 19(2):171–209. DOI: 10.1007/s11036-013-0489-0.

Clarindo, J. P., Castro, J. P. C., and Aguiar, C. D. (2021). Combining fog and cloud computing to support spatial analytics in smart cities. *Journal of Information and Data Management*, 12(4). DOI: 10.5753/jidm.2021.1798.

Clements, P., Garlan, D., Little, R., Nord, R., and Stafford, J. (2003). Documenting software architectures: views and beyond. In *25th International Conference on Software Engineering*, pages 740–741. DOI: 10.1109/ICSE.2003.1201264.

Davoudian, A. and Liu, M. (2020). Big data systems: A software engineering perspective. *ACM Computing Surveys*, 53(5):1–39. DOI: 10.1145/3408314.

Deng, N., Wu, C., Yaseen, A., and Wu, H. (2022). ImmuneData: an integrated data discovery system for immunology data repositories. *Database*, 2022. DOI: 10.1093/database/baac003.

Dong, X. L. and Srivastava, D. (2013). Big data integration. In *Proc. ICDE*, pages 1245–1248. DOI: 10.1109/ICDE.2013.6544914.

Felikson, D., Fenty, I., Hamlington, B., Shiklomanov, A., Blackwood, C., Carroll, M., Croteau, M., David, C., Drushka, K., Duffy, D., et al. (2022). NASA's earth information system: Sea-level change. In *OCEANS 2022, Hampton Roads*, pages 1–8. DOI: 10.1109/OCEANS47191.2022.9977250.

Fernandez, R. C., Pietzuch, P. R., Kreps, J., Narkhede, N., Rao, J., Koshy, J., Lin, D., Riccomini, C., and Wang, G. (2015). Liquid: Unifying nearline and offline big data integration. In *Proc. CIDR*.

Jacobsen, A., de Miranda Azevedo, R., Juty, N., Batista, D., Coles, S., Cornet, R., Courtot, M., Crosas, M., Dumontier, M., Evelo, C. T., et al. (2020). FAIR Principles: Interpretations and implementation considerations. *Data Intelligence*, 2(1-2):10–29. DOI: 10.1162/dint_r_00024.

Jha, A. K., Mithun, S., Sherkhane, U. B., Jaiswar, V., Shi, Z., Kalendralis, P., Kulkarni, C., Dinesh, M. S., Rajamenakshi, R., Sunder, G., Purandare, N., Wee, L., Rangarajan, V., van Soest, J., and Dekker, A. (2022). Implementation of big imaging data pipeline adhering to FAIR principles for federated machine learning in oncology. *IEEE Transactions on Radiation and Plasma Medical Sciences*, 6(2):207–213. DOI: 10.1109/TRPMS.2021.3113860.

Kiran, M., Murphy, P., Monga, I., Dugan, J., and Baveja, S. S. (2015). Lambda architecture for cost-effective batch and speed big data processing. In *Proc. IEEE Big Data*, pages 2785–2792. DOI: 10.1109/BigData.2015.7364082.

Kreps, J. (2014). Questioning the Lambda architecture. Available at https://www.oreilly.com/radar/questioning-the-lambda-architecture/.

Lehmann, J., Schorz, S., Rache, A., Häußermann, T., Rädle, M., and Reichwald, J. (2023). Establishing reliable research data management by integrating measurement devices utilizing intelligent digital twins. *Sensors*, 23(1):468. DOI: 10.3390/s23010468.

Maedche, A., Elshan, E., Höhle, H., Lehrer, C., Recker, J., Sunyaev, A., Sturm, B., and Werth, O. (2024). Open science: Towards greater transparency and openness in science. *Business & Information Systems Engineering*, pages 1–16. DOI: 10.1007/s12599-024-00858-7.

Martinez-Fernandez, S., Medeiros Dos Santos, P. S., Ayala, C. P., Franch, X., and Travassos, G. H. (2015). Aggregating empirical evidence about the benefits and drawbacks of software reference architectures. In *2015 ACM/IEEE International Symposium on Empirical Software Engineering and Measurement (ESEM)*, pages 1–10. DOI: 10.1109/ESEM.2015.7321184.

Martínez-Prieto, M. A., Cuesta, C. E., Arias, M., and Fernández, J. D. (2015). The solid architecture for real-time management of big semantic data. *Future Generation Computer Systems*, 47:62–79. DOI: 10.1016/j.future.2014.10.016.

Medeiros, C. B., Darboux, B. R., Sánchez, J. A., Tenkanen, H., Meneghetti, M. L., Shinwari, Z. K., Montoya, J. C., Smith, I., McCray, A. T., and Vermeir, K. (2020). *IAP input into the UNESCO Open Science Recommendation*. Available at https://www.interacademies.org/sites/default/files/2020-07/Open_Science_0.pdf.

Nadal, S., Herrero, V., Romero, O., Abelló, A., Franch, X., Vansummeren, S., and Valerio, D. (2017). A software reference architecture for semantic-aware big data systems. *Information and Software Technology*, 90:75–92. DOI: 10.1016/j.infsof.2017.06.001.

Nakagawa, E. Y., Antonino, P. O., and Becker, M. (2011). Reference architecture and product line architecture: A subtle but critical difference. In *Proc. ECSA*, pages 207–211. DOI: 10.1007/978-3-642-23798-0_22.

Pana, G. T., Ivanoaica, T., Raportaru, M. C., Baran, V., and Nicolin, A. (2021). Towards the implementation of FAIR principles on an earthquake analysis platform. In *Proc. RoEduNet*, pages 1–4. DOI: 10.1109/RoEduNet54112.2021.9638283.

Pestryakova, S., Vollmers, D., Sherif, M. A., Heindorf, S., Saleem, M., Moussallem, D., and Ngomo, A. C. N. (2022). CovidPubGraph: A FAIR knowledge graph of COVID-19 publications. *Scientific Data*, 9(1):389. DOI: 10.1038/s41597-022-01298-2.

Reisen, M. V., Stokmans, M., Basajja, M., Ong'ayo, A. O., Kirkpatrick, C., and Mons, B. (2020). Towards the tipping point for FAIR implementation. *Data Intelligence*, 2(1-2):264–275. DOI: 10.1162/dint_a_00049.

Rueda-Ruiz, A. J., Ogáyar-Anguita, C. J., Segura-Sánchez, R. J., Béjar-Martos, J. A., and Delgado-Garcia, J. (2022). SPSLiDAR: towards a multi-purpose repository for large scale LiDAR datasets. *International Journal of Geographical Information Science*, 36(5):992–1011. DOI: 10.1080/13658816.2022.2030479.

Rusu, O., Halcu, I., Grigoriu, O., Neculoiu, G., Sandulescu, V., Marinescu, M., and Marinescu, V. (2013). Converting unstructured and semi-structured data into knowledge. In *Proc. RoEduNet*, pages 1–4. DOI: 10.1109/RoEduNet.2013.6511736.

Scannavino, K. R. F., Nakagawa, E. Y., Fabbri, S. C. P. F., and Ferrari, F. C. (2017). *Revisão Sistemática da Literatura em Engenharia de Software: teoria e prática*. Elsevier.

Schwagereit, F., Romacker, M., Richard, F., Trypuz, R., Liener, T., and Roche, O. (2022). FAIR data APIs in the FAIR in vivo data sharing platform. In *CEUR Workshop Proc.*

Sciacca, E., Krokos, M., Bordiu, C., Brandt, C., Vitello, F., Bufano, F., Becciani, U., Raciti, M., Tudisco, G., Riggi, S., Topa, E., Azzi, S., Kyd, B., Mantovani, S., Vettorello, L., Tan, J., Quintana, J., Campos, R., and Pina, N. (2022). Scientific visualization on the cloud: the NEANIAS services towards EOSC integration. *Journal of Grid Computing*, 20(1):7. DOI: 10.1007/s10723-022-09598-y.

Sharma, S. and Mangat, V. (2015). Technology and trends to handle big data: Survey. In *Proc. ICACCT*, pages 266–271. DOI: 10.1109/ACCT.2015.121.

Toulet, A., Michel, F., Bobasheva, A., Menin, A., Dupré, S., Deboin, M.-C., Winckler, M., and Tchechmedjiev, A. (2022). ISSA: generic pipeline, knowledge model and visualization tools to help scientists search and make sense of a scientific archive. In *Proc. ISWC*, pages 660–677. DOI: 10.1007/978-3-031-19433-7_38.

Tsai, C.-W., Lai, C.-F., Chao, H.-C., and Vasilakos, A. V. (2015). Big data analytics: A survey. *Journal of Big Data*, 2:1–32. DOI: 10.1186/s40537-015-0030-3.

Vazquez, P., Hirayama-Shoji, K., Novik, S., Krauss, S., and Rayner, S. (2022). Globally accessible distributed data sharing (GADDS): A decentralized FAIR platform to facilitate data sharing in the life sciences. *Bioinformatics*, 38:3812–3817. DOI: 10.1093/bioinformatics/btac362.

Wilkinson, M. D., Dumontier, M., Aalbersberg, I. J., Appleton, G., Axton, M., Baak, A., Blomberg, N., Boiten, J.-W., Santos, L. B. S., Bourne, P. E., *et al.* (2016). The FAIR Guiding Principles for scientific data management and stewardship. *Scientific Data*, 3(1):1–9. DOI: 10.1038/sdata.2016.18.

Wrembel, R. (2017). Novel Big Data Integration Techniques - What is New. BigNovelTI - Panel Discussion. Available at: https://www.essi.upc.edu/dtim/bignovelti/documents/BigNovelTI-Panel-Robert-Wrembel.pdf.